\documentclass[journal]{IEEEtran}
\usepackage{cite}
\usepackage{ulem}
\usepackage{bm}
\usepackage{graphicx}
\usepackage{amsmath,bm}
\usepackage{array}
\usepackage{mdwmath}
\usepackage{amssymb}
\usepackage{mdwtab}
\usepackage{stfloats}
\usepackage{amsmath,amsthm}
\usepackage{threeparttable}
\usepackage{color}
\usepackage{ulem}
\usepackage[ruled,linesnumbered]{algorithm2e}
\usepackage{multirow}
\usepackage[caption=false,font=normalsize]{subfig}
\usepackage{fancyhdr}

\makeatletter
\newcolumntype{"}{@{\hskip\tabcolsep\vrule width 3pt\hskip\tabcolsep}}
\makeatother

\begin{document}
	\title{\huge A Novel BVH-accelerated Dynamic Ray-Tracing Channel Model for Vehicular Communication Systems}
	\author{Songjiang Yang, \IEEEmembership{Member,~IEEE}, Cheng-Xiang Wang, \IEEEmembership{Fellow,~IEEE}, Yinghua Wang, \\Chen Wang, Jie Huang, \IEEEmembership{Senior Member,~IEEE}, and Yunfei Chen, \IEEEmembership{Fellow,~IEEE}
	\thanks{This work was supported by the National Natural Science Foundation of China (NSFC) under Grants 62401644, the Major Science and Technology Project	of Jiangsu Province under Grant BG2025039, and the Research Fund of National Mobile Communications Research Laboratory, Southeast University, under Grant 2026A05. This work was presented in part at the IEEE Vehicular Technology Conference-Spring 2024 (VTC-Spring'24), Singapore \cite{Wang2024VTC}. (\textit{Corresponding Author: Cheng-Xiang Wang})}
		
	\thanks{S. Yang is with the School of Communication and Information Engineering, Shanghai University, Shanghai 200444, China (e-mail: yangsongjiang@shu.edu.cn).
			
	C.-X. Wang and J. Huang are with the National Mobile Communications Research Laboratory, School of Information Science and Engineering, Southeast University, Nanjing 211189, China, and also with the Purple Mountain Laboratories, Nanjing 211111, China (email: \{chxwang, j\_huang\}@seu.edu.cn).
			
	Y. Wang is with Purple Mountain Laboratories, Nanjing 211111, China (email: wangyinghua@pmlabs.com.cn).
				
	C. Wang is with the Realtek Semiconductor Corporation (Suzhou), Suzhou 215027, China (cecilia\_wang@realsil.com.cn).

	Y. Chen is with the Department of Engineering, University of Durham, Durham DH1 3LE, U.K. (e-mail: yunfei.chen@durham.ac.uk).	} }
	\markboth{IEEE Journal on Selected Areas in Communications,~Vol.~XX, No.~XX, Month~2026}
	{Shell \MakeLowercase{\textit{et al.}}: Bare Demo of IEEEtran.cls for Journals}
	\maketitle

\begin{abstract}
	Accurate and efficient channel modeling is critical for evaluating the sixth-generation (6G) networks, while dynamic ray-tracing (DRT) remains bottlenecked by its reliance on channel coherence time. 
	To address this limitation, this paper proposes a novel bounding volume hierarchy-accelerated DRT (BVH-DRT), which introduces a mechanism shift from a time-driven update in DRT to a spatial-driven update. 
	The proposed channel model clusters complex scatterers using the BVH method as pre-processing to establish a geometric coherence distance guided by transceiver kinematics, so that the ray update intervals in BVH-DRT can be autonomously adjusted to reduce the complexity. 
	Furthermore, the  non-stationary channel impulse responses are generated to evaluate delay-Doppler (DD) domain waveforms, including orthogonal time frequency space (OTFS) and orthogonal delay-Doppler division multiplexing (ODDM). 
	Validations in vehicle-to-vehicle (V2V) scenarios demonstrate that BVH-DRT achieves spatial-temporal consistency with measurements and commercial software while significantly enhancing computational efficiency. 
	Additionally, bit error rate (BER) evaluation highlights the severe impact of high-dynamic channels on waveform performance. 
	The proposed framework is important for designing reliable communications and robust sensing in 6G networks.
\end{abstract}


\IEEEpeerreviewmaketitle

\section{Introduction}
	\normalem
	The sixth-generation (6G) networks are poised to support emerging applications, including integrated sensing and communications (ISAC) and vehicular networks \cite{CXWang2023COMST}. 
	A defining characteristic of these scenarios is the prevalence of high-dynamic channels, which induce severe time and frequency selectivity and non-stationary fading \cite{Liu2026JSAC, SYANG2022TVT}. 
	Consequently, to ensure reliable communications and robust sensing in high-dynamic environments, the evaluation of advanced waveforms depends on channel measurements and channel models capable of accurately capturing these rapid variations~\cite{Gao2025IoTJ, Liu2020JSAC}.
	﻿
		
	To overcome the limitations of conventional orthogonal frequency-division multiplexing (OFDM) in high-dynamic scenarios, advanced waveform designs have emerged featuring delay-Doppler (DD) domain waveforms \cite{Lin2022TWC}. 
	Orthogonal time frequency space (OTFS) and orthogonal delay-Doppler division multiplexing (ODDM) map symbols directly onto the DD domain for wireless transmission \cite{Huang2025TCom}. 
	To use these waveforms, the time and frequency selectivity channels are converted into a more stable DD domain representation, offering robust resilience against severe Doppler shifts \cite{Tong2024TCom}.
	However, existing performance evaluations for these waveforms depend on simplified analytical channel models, i.e., Rayleigh channels~\cite{Li2021TWC}. 
	In addition, practical OTFS and ODDM evaluations have adopted the 3GPP tapped-delay-line (TDL) model and clustered-delay-line (CDL) model to provide reproducible link-level conditions~\cite{Tong2024TCom, Gunturu2021WCNC}. 
	We note that performance evaluations of these waveforms using more comprehensive high-dynamic channel models, such as GBSMs and ray-tracing, remain relatively limited.
	
	Channel modeling methods are generally categorized into stochastic and deterministic channel models \cite{Fan2025JSAC}. 
	For high-dynamic scenarios, the geometry-based stochastic model (GBSM) was widely used due to its low computational complexity and good accuracy~\cite{CXwang2023Tcom, CXWang2022TVT}. 
	GBSMs characterized the channel using statistical distributions for parameters like delay and angular information to describe the relative position of the transmitter (Tx), receiver (Rx), and scatterers~\cite{CXWang2025SCIS}. 
	The geometry-based efficient propagation model for vehicle-to-vehicle communications ($\mathrm{GEMV}^{2}$) was a channel simulator considering different obstruction conditions in vehicular communications and verified by the channel measurements in highway environments~\cite{Boban2014TVT}. 
	In \cite{Yang2020TWC}, the authors proposed and validated a cluster-based vehicular channel model, based on 5.9~GHz measurements in urban scenarios, to characterize the multipath effects caused by large vehicle obstructions in obstructed line-of-sight (OLoS) scenarios.
	In~\cite{Liu2025TWC}, the authors investigated the impact of vehicle obstructions on millimeter-wave vehicular channels through measurements in urban and parking scenarios with root-mean-square (RMS) delay spread and K-factor.
	Moreover,  a stochastic model was proposed to analyze channel non-stationary and consistency.
	In \cite{Huang2024TWC}, the authors proposed a novel three-dimensional (3D) GBSM for 6G millimeter-wave vehicular channels, distinguished static and dynamic clusters to capture the mixed-bouncing propagation.
	Furthermore, the visibility region and the birth-death process were integrated to accurately capture channel non-stationary and consistency for vehicular communications with the continuously arbitrary trajectory.
	However, since GBSMs relied on a set of general environmental parameters, their accuracy for specific, complex propagation environments may not meet the requirements for 6G applications \cite{Ma2023IoTJ}.
	
	The ray-tracing is a deterministic channel modeling to achieve the high accuracy required for 6G~\cite{Zhou2025TAP, Qi2026TCCN}. 
	The ray-tracing operates by using geometrical optics (GO) and the uniform theory of diffraction (UTD) to find all possible pathways between the Tx and Rx, and then computes the electromagnetic field for each path using Maxwell's law \cite{SYang2026TWC}.
	A hybrid sensing and communication channel model for vehicle-to-everything was proposed by combining accurate ray-tracing for moving targets with a fast and object-based stochastic method for environmental clutter~\cite{Liu2024IoTJ}.
	However, the ray-tracing is impractical for real-time channel prediction in dynamic scenarios involving mobile transceivers because of the high computational complexity \cite{Lecci2021TWC}. 
	To address this challenge, dynamic ray-tracing (DRT) was proposed to improve computational efficiency in vehicular communications. 
	The concept of DRT is based on performing a single ray-tracing simulation and then extrapolating the results over a short period, where the channel can be considered stationary \cite{Bilibashi2020EuCap, Bilibashi2023TAP}. 
	A numerical approach to track ray interaction points was developed to further enhance DRT for vehicular communications \cite{Quatresooz2021TVT}.
	Moreover, later work has further studied how to decide the extrapolating period in DRT by using the inter-visibility table in the pre-processing~\cite{ SYANG2025TVT, Hussain2022TAP}.
	﻿In addition, integrating ray-tracing with machine learning (ML) techniques to address the computational challenges in high-dynamic scenarios became a new trend  \cite{Bakirtzis2022TAP, Liu2025AWPL}. 
		
	Addressing the computational limitations of ray-tracing channel model in high-dynamic scenarios, different acceleration techniques have been investigated, including parallel computation, central processing unit (CPU) acceleration, graphics processing unit (GPU) acceleration, and spatial-division methods. 
	In \cite{Tiberi2009TAP}, a parallel ray approximation technique based on CPU acceleration was employed to reduce the computational complexity by assuming a constant number of rays within a wavelength region. With the development of GPU computing, Sionna RT was developed to efficiently simulate vehicular communications in complex propagation environments~\cite{hoydis2023sionna}.
	Moreover, there were various space-division acceleration methods with different structures, like uniform grids, octrees, and k-dimension (KD)-trees. However, these space-division acceleration methods may assign the same object to multiple subspaces, leading to redundant intersection tests~\cite{Meister2021CGF}. 
	On the other hand, the bounding volume hierarchy (BVH) as an object-division method can ensure each object primitive exists in only one leaf node to reduce redundancy for dynamic geometries~\cite{YanICCSNT2017}. 
	
	Existing BVH-based ray-tracing methods mainly focus on improving BVH construction and traversal or reducing unnecessary intersection tests during each complete ray-tracing execution.
	In~\cite{Liu2024AWPL}, the authors proposed a bidirectional ray-tracing method using a BVH, enhanced by a surface area heuristic (SAH) and a novel circumradius criterion, to increase computational efficiency for scattering calculations. 
	In~\cite{Wang2022VTC}, the SAH cost function in ray-tracing was improved by jointly considering the surface area and spatial distance to reduce unnecessary intersection tests.
	In~\cite{Wang2024VTC}, the simplified interval BVH (SIBVH) was developed to separately process static and dynamic objects in the environment and restrict the candidate partitioning interval in the SAH procedure, thereby reducing the repeated BVH construction cost in dynamic scenarios.
	Nevertheless, these methods primarily use BVH as a spatial acceleration structure and do not tell when a complete ray-tracing update needs to be performed during high-dynamic channel simulations. Hence, it is important to combine the current acceleration method with the DRT, but this has not been sufficiently studied for high-dynamic scenarios.
	
	In this paper, we propose a BVH-accelerated DRT (BVH-DRT) framework to compute high-dynamic channels in vehicular communications. 
	More specifically, SIBVH is further developed and incorporated as a key pre-processing component of BVH-DRT. 
	Then, the ray creation or suppression points along the trajectory are found based on the generated bounding boxes to support BVH-DRT compute flexibility and efficiency with adaptive simulation time intervals. 
	Finally, how to evaluate the OTFS and ODDM performances based on the high-dynamic channels from BVH-DRT is investigated. 
	The main contributions and novelties of this paper are summarized as follows.
	\begin{itemize}
		\item
		A novel BVH-DRT framework is proposed to model high-dynamic channels efficiently and accurately.  
		More specifically, the proposed BVH-DRT introduces a paradigm shift from a time-driven update mechanism in DRT to a spatial-driven approach. 
		The SIBVH is used as a pre-processing method to enable this paradigm shift by extracting the size of the bounding volumes based on the intersection test along the trajectory of transceivers. 
		The different sizes of the generated bounding volumes support the BVH-DRT computation flexibility.
		\item 
		The SIBVH is further developed to compute the bounding volume for static and dynamic objects in high-dynamic vehicular scenarios. 
		Static and dynamic objects are processed separately using different leaf-node thresholds and partitioning strategies, while the candidate partitioning interval for the dynamic BVH is restricted using the simplified SAH cost function.
		Furthermore, SIBVH is incorporated and further investigated as a key pre-processing component to analyze the updating interval of the proposed BVH-DRT framework.
		\item
		The performance of ODDM and OTFS under high-dynamic channels derived from BVH-DRT is evaluated. 
		By transforming the channel information of the time-frequency domain into the DD domain, a sparse matrix representation of the channel is utilized. 
		The results validate that the ODDM can enhance robustness and strict orthogonality over the OTFS in 6G high-dynamic vehicular communications.
	\end{itemize}  

	The rest of this paper is organized as follows. 
	In Section~II, the principles of BVH-DRT and SIBVH are investigated.
	In Section~III, how to use the DD channel model from BVH-DRT to evaluate the advanced waveform is shown.
	In Section~IV, results about the accuracy and computational efficiency of the proposed BVH-DRT are shown. 
	Finally, in Section V, our main conclusions are drawn.

\section{BVH-accelerated Dynamic Ray-tracing}
	This section presents the proposed BVH-DRT framework to overcome the computational limitations in complex environments of DRT.  
	The BVH-DRT investigated a fundamental paradigm shift from a time-driven update mechanism in DRT to a spatial-driven approach. 
	Moreover, the novel SIBVH method is incorporated as a key pre-processing method for static and dynamic objects to support the proposed BVH-DRT.
	
	To characterize the high-dynamic channel, the DRT tracks the temporal evolution of all interacting points along the trajectory of transceivers. 
	Initially, a comprehensive ray-tracing simulation is executed to extract the exact ray information under the setting environments. 
	Leveraging the known initial positions, velocities, and accelerations of the transceivers and environmental scatterers, DRT deterministically extrapolates the ray information to compute the channel using analytical formulas. 
	However, the extrapolation is only valid throughout the coherence time concept, during which no major ray creation or suppression should occur, and the underlying topological structure of the ray information remains stationary~\cite{Bilibashi2023TAP}. 
	The actual topological duration of a ray depends heavily on the geometry of the environment. 
	In practice, DRT often requires empirical trial-and-error procedures to optimize the time subdivisions to match actual environment variations. 
	Consequently, relying on an unknown and dynamically fluctuating coherence time restricts the capability of DRT to autonomously and accurately capture the sudden birth or death of multipath components in realistic vehicular networks.
	
\subsection{Spatial-driven DRT Framework}
\begin{figure}[t]
		\centering
		\includegraphics[width=0.85\linewidth]{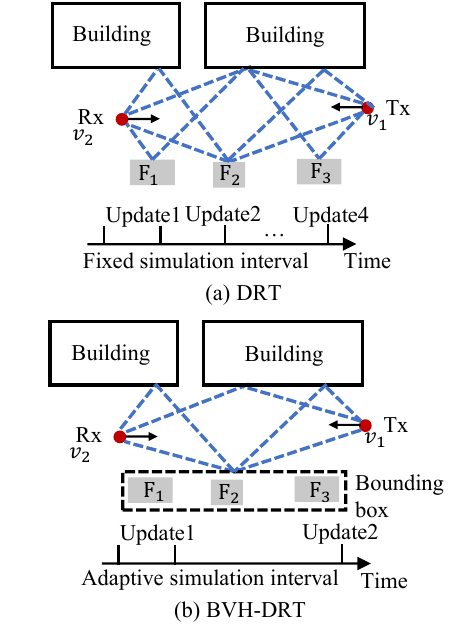}
		\caption{Comparison of  mechanisms between DRT and proposed BVH-DRT.}
		\label{fig:BVH-DRT}
		\vspace{-0.5cm}
	\end{figure}

	To overcome the computational bottlenecks of DRT, the proposed BVH-DRT framework introduces a paradigm shift from a time-driven update mechanism to a spatial-driven approach, heavily leveraging the BVH as an environment pre-processing method.
	The comparison of concepts between the DRT and BVH-DRT is shown in Fig. \ref{fig:BVH-DRT}.
	First, it is important to clearly distinguish two foundational concepts: the channel coherence time and the ray creation and suppression time~$T_\mathrm{cs}$. 
	The coherence time is an intrinsic physical characteristic of the wireless channel, representing the duration over which the multipath remains strictly invariant. 
	The coherence time serves as the fundamental physical assumption in both DRT and BVH-DRT, where the ray information is updated by the analytical extrapolation in this period. 
	In contrast, $T_\mathrm{cs}$ is the setting parameter in BVH-DRT that dictates how often the dynamic simulation updates the ray information by using ray-tracing. 
	Since the BVH method restructures the environmental map during the pre-processing phase, BVH-DRT evaluates channel stability through a geometric coherence distance $D_\mathrm{c}$ retrieved directly from this pre-computed hierarchy, rather than relying blindly on an arbitrary time step. 
	In BVH-DRT, the ray creation and suppression time are deterministically and rapidly computed based on this pre-processed map geometry and the kinematic properties of the vehicles, expressed as
	\begin{equation}
		T_\mathrm{cs} = \frac{D_\mathrm{c}}{|\mathbf{v}|}
		\label{equ:Tcs}
	\end{equation}
	where $\mathbf{v}$ represents the relative velocity vector of the dynamic objects in the environment.
	
	The integration of BVH as a pre-processing step fundamentally accelerates the DRT through two primary mechanisms driven by this mathematical relationship. 
	First, in DRT, the presence of numerous small scatterers with tiny individual lengths $l_\mathrm{s}$ forces the coherence time to become small, which is shown as
	\begin{equation}
		T_\mathrm{c} \approx \frac{l_\mathrm{s}}{ |\mathbf{v}|}.
		\label{equ:Tc}
	\end{equation}	
	This shrinks the extrapolation intervals in DRT, leading to excessive computational overhead as the execution time of the comprehensive ray-tracing increases. 
	BVH resolves this problem during the pre-processing by clustering these small objects into a unified bounding volume with an effective macroscopic length, $L_\mathrm{BV}$. 
	The multipath birth and death mechanism for these bounding volumes will remain highly similar to the original small scatterers.  	
	By expanding the coherence distance to the bounds of the pre-computed bounding volume ($D_\mathrm{c} = L_\mathrm{BV}$), the ray creation and suppression time can be set significantly larger than the coherence time of the individual small objects, drastically reducing the computational burden.
	Second, DRT lacks a robust method for determining the optimal extrapolation interval, often relying on trial and error. 
	By employing fixed bounding boxes with relatively uniform spatial bounding, the BVH framework provides a structured geometric baseline. 
	This geometric structure allows BVH-DRT to adaptively adjust the extrapolation interval based on the time it takes a dynamic object to traverse a specific bounding box.
	As the vehicle moves through different spatial clusters in the pre-processed map, the BVH-DRT dynamically recomputes ray creation and suppression time for each specific BVH node. 
	This adaptive adjustment successfully balances high-accuracy temporal tracking with maximized computational efficiency. 
	
\subsection{SIBVH}
	BVH methods utilize a binary tree-based acceleration structure comprising two primary phases: construction and traversal. 
	The construction phase focuses on creating a binary tree dependent on the current geometric information of the environment. 
	To employ BVH effectively in high-dynamic scenarios, minimizing the overlap of triangular facets within bounding volumes is the key point. 
	To ensure robustness in these environments, it is important to decouple the modeling of static and dynamic objects based on their distinct kinematic properties. 
	Static objects exhibit time-invariant spatial attributes, whereas dynamic objects are changed across each snapshot. 
	Consequently, treating these dynamic objects independently is essential to accommodate their specific update requirements and reduce tree-building complexity.

	\begin{figure}[b]
		\vspace{-0.5cm}
		\centering
		\includegraphics[width=0.85\linewidth]{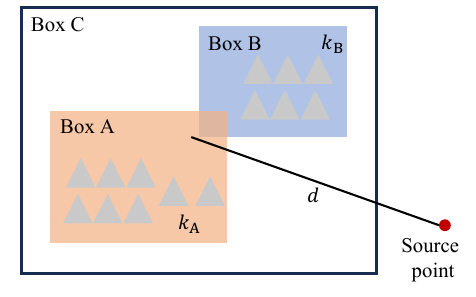}
		\caption{Uneven distribution and the invalid overlap problems for BVH.}
		\label{fig:bvh}
	\end{figure}
	
	In the BVH, SAH methods are widely used to analyze the optimization of BVH construction because the SAH can find the most suitable partitioning strategy.
	The improved SAH algorithm is proposed by considering the dual effects of surface area and spatial distance to find the partitioning strategy accurately and efficiently. 
	Accordingly, the uneven distribution of left and right child nodes and the invalid overlap of bounding boxes can be overcome by improved SAH. 
	In Fig.~\ref{fig:bvh}, the parent node bounding box is denoted as C. 
	The left and right node bounding box denote as A and B, respectively. 
	Furthermore, ${k}_\text{A}$ and ${k}_\text{B}$ are denoted as the number of triangular facets in bounding boxes A and B,~respectively. 
	The centroid-based spatial-distance term between the ray source and the candidate child bounding boxes is represented by $d$. 
	Hence, the cost function of the SAH can be expressed as \cite{Wang2024VTC}
	\begin{equation}
		\label{equation: old SAH}
		\eta(\mathrm{A},\mathrm{B})={\alpha}\left(\frac{{S}(\mathrm{A})}{{S}(\mathrm{C})}{k}_{\mathrm{A}}{\eta}_{i}+\frac{{S}(\mathrm{B})}{{S}(\mathrm{C})}{k}_{\mathrm{B}}{\eta}_{i}\right)+(1-\alpha){d}^2+{\eta}_{\mathrm{tr}}
	\end{equation}
	where $\eta_{\mathrm{tr}}$ represents BVH tree structure construction cost, ${\eta}_{i}$ represents intersection cost, ${\alpha}$ is the weight coefficient between the surface area of the bounding volume and spatial distance, and $S(\mathrm{A})$, $S(\mathrm{B})$, and $S(\mathrm{C})$ represent surface areas of bounding boxes $\mathrm{A}$, $\mathrm{B}$, and C, respectively. 
	The value of the weight coefficient ${\alpha}$ is in the range of [0,1].
	
	Since the static objects are not changing after the initial snapshot, the BVH for static objects is constructed once and reused, to reduce the construction cost over subsequent snapshots.  Using the SAH, Algorithm~1 recursively partitions the triangular facets until each leaf contains facets no more than  the leaf-node threshold. A smaller leaf-node threshold $N_t$ creates more leaf nodes and a deeper tree, generally increasing the construction time.
	Algorithm~1 provides the pseudocode for the construction of the BVH tree, and outlines the BVH process for static objects. 
	The algorithm iteratively partitions the objects into left and right child nodes utilizing the SAH, generating leaf nodes once the number of enclosed objects falls below a predefined threshold. It can be seen from Algorithm 1 that the size of the threshold affects the depth and efficiency of tree-building.
	
	\begin{algorithm}[t]   
		\SetAlgoLined   
		\SetKwFunction{BuildBVH}{BuildBVH}      
		\KwIn{Object list $objects$, Threshold $N_t$}   
		\KwOut{BVH tree internal node}      
		\BlankLine   
		\SetKwProg{Fn}{Function}{:}{}   
		\Fn{\BuildBVH{$objects$}}{     
			\If{the number of objects does not exceed $N_t$}{          
				create leaf node containing these objects;         
				\Return leaf node\;
			}     
			Calculate the bounding volumes for all $objects$\;
			Split $objects$ into two sets $leftObjects$, $rightObjects$ by SAH algorithm\; 
			\BlankLine}
		\textbf{End Function}\; 
		$leftNode \leftarrow$ \BuildBVH{$leftObjects$}\;     
		$rightNode \leftarrow$ \BuildBVH{$rightObjects$}\;     
		\Return internal node with $leftNode$ and $rightNode$ as children.       
		\caption{BVH Construction} 
	\end{algorithm}
	
	If the BVH executed all time in dynamic acceleration, these updated steps will cause considerable redundancy during the bounding boxes partitioning process.
	To overcome these challenges, sacrificing a slight partitioning balance for enhanced computational efficiency becomes a meaningful trade-off of real-time BVH tree construction for high-dynamic scenarios.
	According to \eqref{equation: old SAH}, to minimize $\eta(\mathrm{A},\mathrm{B})$, it can be  transferred to minimize $\frac{{S}(\text{A})}{{S}(\text{C})}{k}_\text{A}$ and $\frac{{S}(\text{B})}{{S}(\text{C})}{k}_\text{B}$. 
	This implies the necessity of containing a substantial number of triangular facets within a confined spatial region and ensuring a balanced distribution of triangular facets between different child nodes.
	For all candidate partitions of the same parent node, $S(\mathrm{C})$, $\eta_i$, and $\eta_{\mathrm{tr}}$ are invariant. Hence, subtracting $\eta_{\mathrm{tr}}$ and normalizing the remaining terms by the common positive factor do not affect the relative ranking of the candidate partitions. The corresponding effective	weighting coefficient is
	\begin{equation}
		\alpha_{\mathrm{eff}} =	\frac{\alpha\eta_i/S(\mathrm{C})}		{\alpha\eta_i/S(\mathrm{C})+(1-\alpha)}.
	\end{equation}
	Accordingly, the normalized cost function can be expressed as
	\begin{equation}
		\label{equation: simplified SAH}
		\eta(\mathrm{A},\mathrm{B})	= \alpha_{\mathrm{eff}} \left[S(\mathrm{A})k_{\mathrm{A}}		+S(\mathrm{B})k_{\mathrm{B}}\right]		+(1-\alpha_{\mathrm{eff}})d^2.
	\end{equation}
	For notational simplicity, $\alpha_{\mathrm{eff}}$ is denoted by $\alpha$.
	This simplification preserves the relative cost comparison among the candidate partitions of the same parent node.
	Thus, the simplified equation aids in reducing computational workload and mitigates a portion of the tree-building time.
	
\begin{figure}[t]
		\centering
		\includegraphics[width=0.85\linewidth]{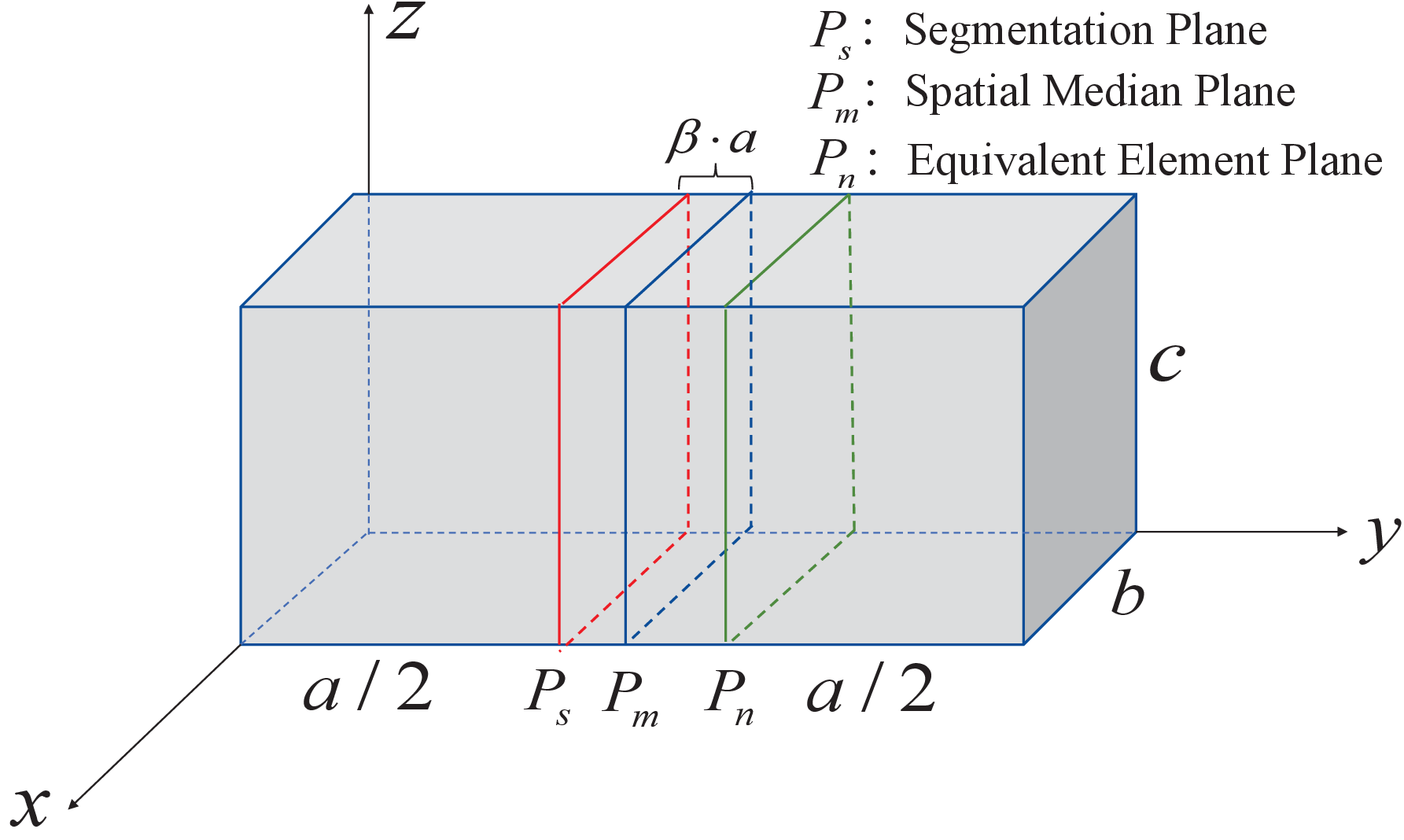}
		\caption{SIBVH partitioning strategy.}
		\label{fig:BVHpar}
		\vspace{-0.5cm}
	\end{figure}

	The SAH algorithm necessitates the computation of $\eta(\mathrm{A},\mathrm{B})$ for every bounding box, and subsequently selects the node with the minimum $\eta(\mathrm{A},\mathrm{B})$ among all candidates.  
	Since the optimal splitting plane within a specific sub-interval can be estimated beforehand, there is no need to compute $\eta(\mathrm{A},\mathrm{B})$ for other sub-intervals.
	This optimization leads to a reduction in the computational cost of BVH construction. 
	In Fig.~\ref{fig:BVHpar}, the segmentation plane, spatial median plane, and equivalent element plane are set as ${P}_{s}$, ${P}_{m}$, and ${P}_{n}$, respectively. 
	The distance from ${P}_{s}$ to ${P}_{m}$ is computed by $\beta\cdot a$, where $\beta$ is the weight to scale the length and $a$ is the length of the node space. 
	The value of ${\beta}$ is in the range of [0,1].
	Then, the original $S(\text{A})$ and $S(\text{B})$ can be denoted as 
	\begin{equation}
		\label{simplified SAH}
		\begin{aligned}
			S(\text{A}) &=2\left[ ba(1/2-\beta) +ca(1/2-\beta)+bc\right]\\
			&=ab+ac+2bc-2\beta(ab+ac)
		\end{aligned}
	\end{equation}
	\begin{equation}
		\label{simplified SAH}
		\begin{aligned}
			S(\text{B}) &=2\left[ ba(1/2+\beta)+ca(1/2+\beta)+bc\right]\\
			&=ab+ac+2bc+2\beta(ab+ac)
		\end{aligned}
	\end{equation}
	where $a$, $b$, and $c$ are the length, width, and height of the node space range, respectively.
	
	The number of triangular facets located on the left and right sides of ${P}_{m}$ are  ${N}_{1}$ and ${N}_{2}$, respectively. 
	In addition, the number of triangular facets between ${P}_{s}$ and ${P}_{m}$ is ${N}_{0}$. 
	Hence, the number of facets in the left and right child nodes are ${k}_\text{A}$=${N}_{1}$-${N}_{0}$ and ${k}_\text{B}$=${N}_{2}$+${N}_{0}$, respectively. 
	Let $S=ab+ac+2bc$ and $m=ab+ac$, so \eqref{equation: simplified SAH} can be expressed as
	\begin{equation}
		\label{equation: our SAH}
		\begin{aligned}
			\eta(\beta)&={\alpha}\left[(S\!-\!2m\beta)({N}_{1}\!-\!{N}_{0})\!+\!(S\!+\!2m\beta)({N}_{2}\!+\!{N}_{0})\right]\\
			&\;\;\;\;\;+(1-\alpha){d}^{2}\\
			&={\alpha}\left[ S({N}_{1}+{N}_{2})+2m({N}_{2}-{N}_{1})\beta+4m\beta{N}_{0}\right] \\
			&\;\;\;\;\;+(1-\alpha){d}^{2}.
		\end{aligned}
	\end{equation}
	
	The plane with the minimum value of $\eta(\beta)$ is the best segmentation plane. 
	To obtain a tractable candidate interval, the triangular-facet centroids along the selected partitioning axis are locally approximated as uniformly distributed. Under this approximation, the number of facets between $P_{\mathrm{s}}$ and $P_{\mathrm{m}}$ can be expressed as $N_0 \approx k|\beta|$. Hence, \eqref{equation: our SAH} can be expressed as a function of $\beta$ as
	\begin{equation}
		\label{equation: f function}
		\eta(\beta)\!\approx\!{\alpha}\left[ S({N}_{1}\!+\!{N}_{2})\!+\!2m({N}_{2}\!-\!{N}_{1})\beta\!+\!4km\beta|\beta|\right]\!+\!(1\!-\!\alpha){d}^{2}.
	\end{equation}
	
	The surface-area and facet-occupancy terms in \eqref{equation: f function} provide a theoretical guide for determining the reduced candidate interval. The relative values of $N_1$ and $N_2$ determine the position of the equivalent-element plane $P_{\mathrm{n}}$ relative to the spatial median plane $P_{\mathrm{m}}$:
	\begin{itemize}
		\item ${N}_{1}$$\textless$${N}_{2}$: The closer to ${P}_\mathrm{m}$, the smaller the value $\eta(\beta)$ is.
		\item ${N}_{1}$$\textgreater$${N}_{2}$: The minimum $\eta(\beta)$ is between ${P}_\mathrm{n}$ and ${P}_\mathrm{m}$.
		\item ${N}_{1}$=${N}_{2}$: When ${P}_\mathrm{s}$ is equal to ${P}_\mathrm{m}$, $\eta(\beta)$ is minimum.
	\end{itemize}
	Based on the cost function, SIBVH restricts the evaluation to a reduced candidate interval determined by the spatial median plane and the equivalent-element plane. 
	The candidate positions within this interval are evaluated using the simplified cost function, and the one with the minimum evaluated cost is selected. Therefore, the proposed partitioning strategy is a theoretically motivated SAH designed to reduce the repeated reconstruction cost of the dynamic BVH.
	
\begin{figure}[t]
		\centering
		\includegraphics[width=0.85\linewidth]{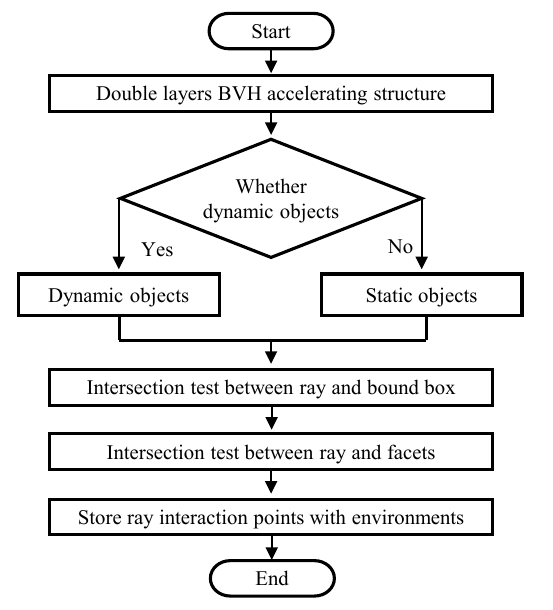}
		\caption{The flowchart of SIBVH pre-processing.}
		\label{fig:SIBVH}
		\vspace{-0.5cm}
	\end{figure}

	The flowchart of the SIBVH pre-processing in the BVH-DRT framework is illustrated in Fig.~\ref{fig:SIBVH}.
	Specifically, the SIBVH constructs an acceleration structure by independently applying the BVH to static objects and the optimized partitioning strategy to dynamic objects. 
	The intersection tests, progressing from bounding boxes down to triangular facets, are executed to accurately extract and store the precise ray interaction points. 
	
\subsection{Extraction of Coherence Distance}
	While the SIBVH algorithm efficiently processes the high-dynamic environmental information, extracting the geometric coherence distance during the dynamic simulation requires a robust traversal and intersection mechanism. 
	When a ray is launched from the Tx, it traverses the pre-constructed SIBVH tree by performing rapid ray-box intersection tests, moving from the root node down to the leaf nodes. 
	Once a valid path interaction point, such as a reflection or diffraction on a specific triangular facet, is confirmed, the minimum BVH leaf node with the specific facet is immediately identified. 
	The framework extracts the spatial dimensions of this localized leaf node. 
	The minimum dimension (length, width, or height) of this bounding box along the trajectory of the dynamic entity is extracted as the effective length, $L_\mathrm{BV}$.
	Consequently, the geometric coherence distance for that specific interaction is dynamically assigned as $D_\mathrm{c} = L_\mathrm{BV}$. 
	Once $D_\mathrm{c}$ is extracted from the leaf node, it is directly substituted into \eqref{equ:Tcs} to compute the exact ray creation and suppression time for the specific propagation path. 
	This targeted extraction ensures that the extrapolation interval in BVH-DRT is adaptively updated based on the localized geometric complexity of the specific scatterers the ray is currently interacting with.
	
	The proposed framework also supports non-uniform motion, such as acceleration and braking. The position and velocity of each dynamic object are updated at every snapshot according to its prescribed trajectory. The static part of the SIBVH remains unchanged and is reused throughout the simulation, whereas the dynamic part of the SIBVH is reconstructed using the updated geometry during each adaptive ray-tracing update, where $T_{\mathrm{cs}}$ is recalculated based on the current relative motion state and the extracted coherence distance $D_{\mathrm{c}}$. Therefore, acceleration or braking does not require reconstructing the complete BVH for the entire environment.
	
\begin{figure}[t]
		\centering
		\includegraphics[width=0.85\linewidth]{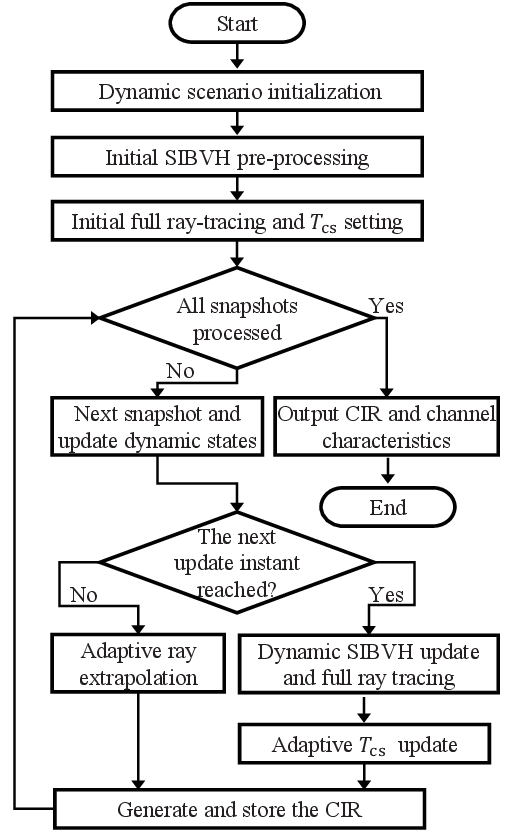}
		\caption{The flowchart of the proposed BVH-DRT framework.}
		\label{fig:flowchart}
		\vspace{-0.5cm}
	\end{figure}

	The overall workflow of the proposed BVH-DRT method is illustrated in Fig.~\ref{fig:flowchart}. At the initial snapshot, the dynamic scenario is initialized, the SIBVH pre-processing is completed, and a full ray-tracing calculation is performed with the initial $T_{\mathrm{cs}}$ setting. For each subsequent snapshot, the positions and motion states of the dynamic objects are updated, and the algorithm checks whether the next adaptively determined ray-update instant has been reached. If not, the current ray information is obtained through analytical ray extrapolation. Otherwise, the dynamic part of the SIBVH is updated, a full ray-tracing calculation is performed, and $T_{\mathrm{cs}}$ is recalculated based on the updated propagation paths as well as the current relative motion state. The recalculated $T_{\mathrm{cs}}$ determines the next ray-update instant. After either ray extrapolation or full ray-tracing, the CIR at the current snapshot is generated and stored. This procedure is repeated until all snapshots have been processed, after which the CIRs and channel characteristics are generated as outputs. The detailed SIBVH pre-processing procedure is presented in Fig.~\ref{fig:SIBVH}. Since the underlying ray-tracing calculation is executed only at the adaptively determined update instants, the proposed BVH-DRT framework can be applied to different ray-tracing platforms.
		
\section{Application to Advanced Waveform Evaluation in Delay-Doppler Domain}
	To demonstrate the practical utility of the proposed BVH-DRT, this section applies the generated high-dynamic channels to evaluate the performance of advanced waveforms.
	Traditional evaluation methods often rely on simplified or purely stochastic models, i.e., Rayleigh fading, which fail to capture the severe, non-stationary geometric blockages inherent in vehicular networks. 
	Since advanced 6G waveforms, such as OTFS and ODDM, map symbols directly onto the DD domain to combat high-dynamic channels, their performance is extremely sensitive to the accuracy of the underlying DD domain channel characteristics~\cite{Lin2022TWC}. 
	By utilizing the BVH-DRT proposed in Section~II, the temporal evolution of the multipath components can be accurately extracted. 
	
	Specifically, the BVH-DRT outputs the instantaneous path amplitude, delay, and Doppler shift for every ray path at any given time $t$. 
	These accurate outputs serve as the direct inputs to construct the non-stationary channel impulse response, which is characterized at time $t$ and carrier frequency $f_\mathrm{c}$ as
	\begin{equation}\label{eq:CIR}
		h(t,f_\mathrm{c})=\sum_{i=1}^I g_i e^{j2 \pi\left(f_\mathrm{c}t+f_{\mathrm{D},i}t-f_\mathrm{c}\tau_i\right)} \delta\left(t-\tau_i\right)
	\end{equation}
	where $I$ is the number of paths, $g_i$ is the path amplitude of the $i$-th path  ($ i \in \{1, 2,...,I\} $), $ f_{\mathrm{D},i} $ is the Doppler shift of the $i$-th path, $ \tau_{i}$ is the delay of the $i$-th path, and $ \delta(\cdot) $ is the Dirac delta function.
	
	To map the high-dynamic channel onto the discrete DD domain channel, the delay and Doppler parameters from $h(t,f_\mathrm{c})$ need to be discretized onto the DD grid.
	For a transmission frame consisting of $P$ time symbols and $Q$ frequency subcarriers, the total frame duration is $T_\mathrm{f} =PT_\mathrm{s}$ and the total system bandwidth is $B = Q \Delta f$, where $T_\mathrm{s}$ and $\Delta f$ denote the symbol duration and subcarrier spacing, respectively. 
	Hence, the resolution of the delay bin $\Delta \tau$ and the Doppler bin $\Delta \nu$ are expressed as
	\begin{equation}
		\Delta \tau = \frac{1}{B} = \frac{1}{Q \Delta f}
	\end{equation}
	\begin{equation}
		 \quad \Delta \nu = \frac{1}{T_\mathrm{f}} = \frac{1}{P T_\mathrm{s}}.
	\end{equation}
	Consequently, the continuous propagation delay $\tau_i$ and Doppler shift $f_{\mathrm{D},i}$ of the $i$-th physical path tracked by the BVH-DRT are mapped to their corresponding discrete delay index $l_i$ and Doppler index $k_i$ on the DD grid according to
	\begin{equation}
		l_i = \lfloor \frac{\tau_i}{\Delta \tau} \rceil
	\end{equation}
	\begin{equation} 
		\quad k_i = \lfloor \frac{f_{\mathrm{D},i}}{\Delta \nu} \rceil
	\end{equation}
	where $\lfloor \cdot \rceil$ denotes the nearest integer rounding operator. 
	By applying these respective waveform basis transformations, the discrete DD domain channel matrix $\mathbf{H}_{\text{eq}} \in \mathbb{C}^{PQ \times PQ}$ is formulated as the linear superposition of all valid paths
		\begin{equation}
		\mathbf{H}_{\text{eq}} = \sum_{i=1}^{I} g_i e^{-j 2\pi f_\mathrm{c} \tau_i} \mathbf{\Psi}(l_i, k_i)
	\end{equation}
	where $g_i e^{-j 2\pi f_\mathrm{c} \tau_i}$ incorporates the equivalent baseband phase shift of the $i$-th ray from the BVH-DRT. 
	The matrix $\mathbf{\Psi}(l_i, k_i) \in \mathbb{C}^{PQ \times PQ}$ represents the DD domain shift operator, which inherently performs the cyclic permutations and phase rotations dictated by the $i$-th path's indices $(l_i, k_i)$.

	Let $\mathbf{x} \in \mathbb{C}^{PQ \times 1}$ denote the vectorized transmitted symbols mapped onto the DD domain. 
	After pulse shaping, propagation through the DD channel, and matched filtering at the Rx, the discrete DD-domain received signal $\mathbf{y} \in \mathbb{C}^{MN \times 1}$ is uniformly modeled as
	\begin{equation}
		\mathbf{y} = \mathbf{H}_{\text{eq}} \mathbf{x} + \mathbf{z}
	\end{equation}
	where $\mathbf{z}$ represents the additive white Gaussian noise. 
	The structural differences in $\mathbf{H}_{\text{eq}}$ between ODDM and OTFS stem from their respective basis orthogonalities and pulse shaping characteristics.
	To decouple the effects of sub-optimal receiver designs and the fundamental capabilities of the waveforms, we employ an exact linear minimum mean square error (LMMSE) equalizer. 
	Based on \cite{Gao2025IoTJ}, the optimally estimated symbol vector is given by
	\begin{equation}
		\hat{\mathbf{x}}_{\text{LMMSE}} = \left( \mathbf{H}_{\text{eq}}^\text{H} \mathbf{H}_{\text{eq}} + \frac{\sigma_z^2}{P_{\text{T}}} \mathbf{I}_{MN} \right)^{-1} \mathbf{H}_{\text{eq}}^\text{H} \mathbf{y}
	\end{equation}
	where $P_{\text{T}}$ is the average transmission power. 
	The perfect channel state information is assumed because $\mathbf{H}_{\mathrm{eq}}$ is constructed from the BVH-DRT results to keep the comparison of the different types of modulation fair. All valid ray-path contributions are retained, while DD-domain bins without valid paths remain zero. Hence, no additional power-based tap-selection threshold is applied.
	Since evaluating the exact LMMSE over a large-scale grid conventionally leads to high computational complexity, we adopt an equivalent sparse linear solver architecture. 
	By exploiting the inherent sparsity of the DD domain channel, this framework entirely bypasses explicit dense matrix inversions while preserving strict mathematical equivalence to the ideal LMMSE solution.
	
	\section{Results and Analysis}
\begin{figure}[t]
		\centering
		\includegraphics[width=0.85\linewidth]{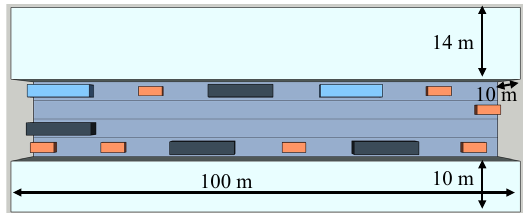}
		\caption{Urban vehicular communications scenario.}
		\label{fig:scenario}
		\vspace{-0.5cm}
	\end{figure}

	To validate the proposed BVH-DRT, a typical urban vehicular communications scenario is constructed, as illustrated in Fig. \ref{fig:scenario}. 
	The simulation area is 100~m in length and 40~m in width, comprising 4 m wide driving lanes and densely distributed parking spaces along both sides. 
	To simplify the environment while retaining the primary static scatterers, the spacing between buildings is omitted. 
	Specifically, the size of upper building is 100~m~$\times$~14~m~$\times$~10~m (length $\times$ width $\times$ height), and the size of lower building is 100~m~$\times$~10~m~$\times$~10~m. 
	Both buildings are enclosed by concrete walls and glass windows. 
	This layout effectively models real-world urban street canyons~\cite{AbbasIJAP2015} to evaluate vehicle-to-vehicle (V2V)  propagation scenarios. 
	According to ITU-R P.2040~\cite{ITU2023}, the electromagnetic properties of the materials within the environment, including concrete, ground, metal, and glass, are summarized in Table~\ref{tab:material}. 
	
	To ensure the environment in simulation is close to the real world, three different vehicle types with metal main body and glass windows are used following the 3GPP TR 37.885 standard~\cite{3GPP2019Study}, as detailed in Table~\ref{tab:vehicular}. 
	The communication link is established between a Tx mounted on a bus and a Rx positioned on a small car, traveling in opposite lanes. 
	The common simulation settings in this work are summarized in Table~\ref{tab:simulation_settings}. Based on this environment topology and the vehicular speeds, the initial ray creation and suppression time for the DRT and BVH-DRT is set to 0.3 s. Furthermore, a maximum reflection order of 3 is configured, which provides sufficient multipath components to accurately capture the dominant propagation characteristics within the urban environment. In particular, the initial positions and trajectories of the Tx and Rx are kept identical in the two cases to ensure a fair comparison. In Case 1, there are no moving scatterers, where the effect of obstructed vehicles is not considered. In Case~2, an additional moving obstructed vehicle is presented, where the obstructed vehicle is in the same lane as the Tx. The initial center of the moving bus is located at (26.5,16,1.5) m, and it travels at 15 m/s in the same lane and direction as the Tx. Except for this additional moving vehicle, all other simulation settings remain the same for the two cases.
	
	\begin{table}[t]
		\centering
		\caption{The parameters of the different materials.}
		\renewcommand{\arraystretch}{1.2}
		\begin{tabular}{|c|c|c|c|}
			\hline
			Material & Permittivity & Conductivity (S/m) & Thickness (m) \\
			\hline
			Concrete & 5.31 & 0.12 & 0.3 \\
			\hline
			Ground & 3 & 0.087 & 1 \\
			\hline
			Metal & 1 & $10^{7}$ & 0.08 \\
			\hline
			Glass & 6.27 & 0.02931 & 0.03 \\
			\hline
		\end{tabular}
		\label{tab:material}
	\end{table}

	\begin{table}[t]
		\centering
		\caption{The parameters of the different vehicle types.}
		\renewcommand{\arraystretch}{1.2}
		\begin{tabular}{|c|c|c|c|}
			\hline
			Vehicle type & Car (brown) & Bus (blue) & Truck (black) \\
			\hline
			Length (m) & 5 & 13 & 13.5 \\
			\hline
			Width (m) & 2 & 2.6 & 2.6 \\
			\hline
			Height (m) & 1.6 & 3 & 3.5 \\
			\hline
		\end{tabular}
		\label{tab:vehicular}
		\vspace{-0.5cm}
	\end{table}
	
	\begin{table}[t]
		\centering
		\caption{Common simulation settings for Cases 1 and 2~\cite{Boban2014TVT, 3GPP2019Study}.}
		\label{tab:simulation_settings}
		\renewcommand{\arraystretch}{1.2}
		\begin{tabular}{|c|c|}
			\hline
			Parameters & Value \\
			\hline
			Transmission power & 0 dBm\\
			\hline
			Tx/Rx antenna & Omnidirectional (0~dBi) \\
			\hline
			Carrier frequency & 5.9~GHz \\
			\hline
			Initial $T_{\mathrm{cs}}$ & 0.3~s \\
			\hline
			Maximum reflection order & 3 \\
			\hline
			Tx initial position & $(13.5,16,3.7)$~m \\
			\hline
			Tx motion & 15~m/s along the $+x$ direction \\
			\hline
			Rx initial position & $(95,20,1.6)$~m \\
			\hline
			Rx motion & 10~m/s along the $-x$ direction \\
			\hline
		\end{tabular}
	\vspace{-0.2cm}
	\end{table}
	
	\subsection{Channel Characteristics Analysis}
\begin{figure}[t]
		\centering
		\includegraphics[width=0.95\linewidth]{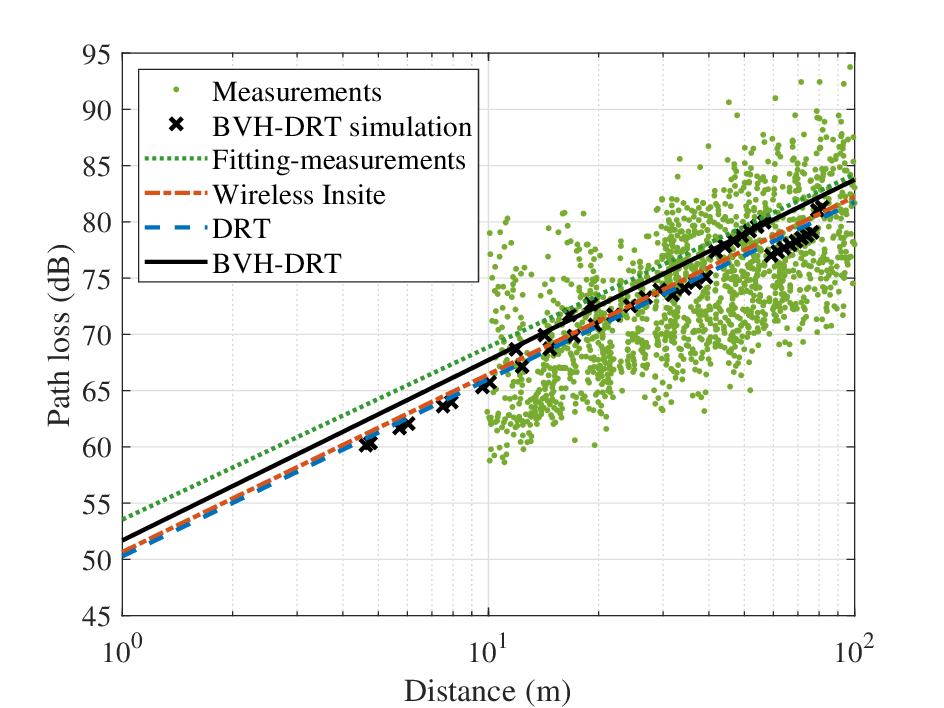}
		\caption{Comparisons of pathlosses with obstructed vehicles among measurement \cite{AbbasIJAP2015}, Wireless Insite, DRT, and BVH-DRT.}
		\label{fig:pathloss}
		\vspace{-0.5cm}
	\end{figure}

	Fig. \ref{fig:pathloss} illustrates the path loss comparison among measurements \cite{AbbasIJAP2015}, Wireless Insite \cite{WI2025}, DRT \cite{Bilibashi2023TAP}, and the proposed BVH-DRT. 
	The measurement data captures a highly complex urban propagation environment characterized by multi-story buildings, streetside scatterers, e.g., trees and signs, and dynamic vehicular traffic moving in both convoy and opposite directions. 
	To quantify the propagation characteristics across these channel models, the log-distance path loss model~\cite{Boban2014TVT} is adopted as
	\begin{equation}
		\label{equ:log-dist}
		P_\mathrm{L}(d) = P_\mathrm{L}(d_0) + 10\gamma\log\left(\frac{d}{d_{0}}\right)+ \sigma
	\end{equation}
	where $\gamma$ is the path loss exponent (PLE), $d_{0}$ is the reference distance, and $\sigma$ accounts for shadowing effects. 
	As summarized in Table \ref{tab:pathlosspara}, the parameters of path loss model for BVH-DRT exhibit excellent agreement with the measurement data. 
	The fitting line of the BVH-DRT is in good agreement with the fitting line of measurement result. 
	The results from Wireless Insite also align tightly with the measurements, justifying its use as the benchmark for the remaining analyses in this paper. 
	\begin{table}[t]
		\centering
		\caption{The path loss model parameters for vehicles with moving scatterers.}
		\renewcommand{\arraystretch}{1.2}
		\begin{tabular}{|c|c|c|}
			\hline
			&$\gamma$ & $ \sigma $ \\
			\hline
			Measurements & 1.505 & 4.6270 \\
			\hline
			Wireless Insite & 1.579 & 1.0518 \\
			\hline
			DRT & 1.574 & 1.0097 \\
			\hline
			BVH-DRT & 1.603 & 1.0984 \\
			\hline
		\end{tabular}
		\label{tab:pathlosspara}
		\vspace{-0.3cm}
	\end{table}
		
\begin{figure}[t]
		\centering
		\subfloat[Scenario without obstructed vehicles]{\includegraphics[width=0.45\textwidth]{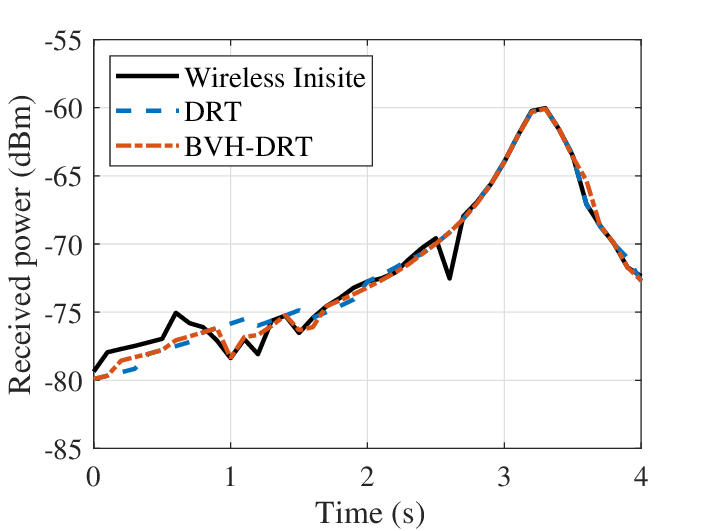}}
		\hfil
		\subfloat[Scenario with obstructed vehicles]{\includegraphics[width=0.45\textwidth]{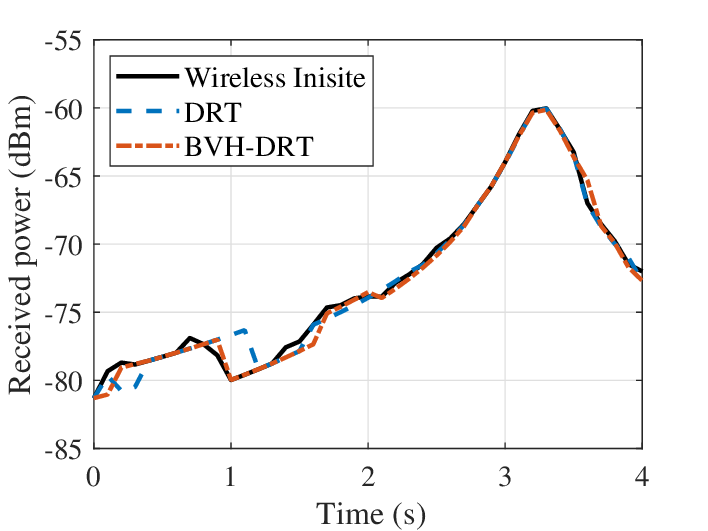}}
		\hfil
		\caption{Comparisons of received power among wireless insite, DRT, and BVH-DRT with/without obstructed vehicles: (a) scenario without obstructed vehicles and (b) scenario with obstructed vehicles.}
		\label{fig:Receivedpower}
		\vspace{-0.5cm}
	\end{figure}

	Fig. \ref{fig:Receivedpower} illustrates the received power obtained via Wireless Insite, DRT, and the proposed BVH-DRT. 
	Across both methodologies, BVH-DRT consistently demonstrates superior accuracy over the DRT.
	In Fig. \ref{fig:Receivedpower}(a), the simulation results of the BVH-DRT exhibit high consistency with both Wireless Insite and DRT, confirming its applicability and accuracy in complex urban scenarios.
	Specifically, compared to the Wireless InSite, the root mean square error (RMSE) of the proposed BVH-DRT achieves 0.8588 dB, outperforming the RMSE of DRT. 
	This accuracy enhancement is particularly evident during the 0.2–0.3 s and 0.9–1.2 s intervals, where this improvement stems from a more reasonable ray creation and suppression time. 
	The BVH-DRT more reasonably manages ray generation and suppression, promptly capturing new rays while discarding outdated multipath components.

	The advantages of the proposed BVH-DRT become more significant in Case 2, as shown in Fig. \ref{fig:Receivedpower}(b). 
	When comparing the two cases, the performance of BVH-DRT actually improves in the high-dynamic environment of Case~2, yielding a RMSE of 0.5558 dB against Wireless Insite, as shown in Table~\ref{tab:recepower}. 
	This performance, particularly during the 0.3–0.5~s and 0.8–1.1~s, is attributed to the adaptive ray creation and suppression time of BVH-DRT. 
	The adaptive mechanism swiftly reacts to sudden environmental alterations and initiates new ray-tracing iterations in time. 
	More specifically, in Case 2, the moving obstructed vehicle causes more frequent changes in the environment, thereby triggering more frequent adaptive updates. The shorter extrapolation duration between successive updates reduces the accumulation of ray-extrapolation errors. Nevertheless, the lower RMSE observed in Case 2 is specific to the considered environments and does not imply that obstructed environments generally achieve lower RMSE values.
	\begin{table}[t]
		\centering
		\caption{The RMSE of received power for BVH-DRT and DRT compared with Wireless Insite.}
		\renewcommand{\arraystretch}{1.2}
		\begin{tabular}{|c|c|c|}
			\hline
			&BVH-DRT &DRT \\
			\hline
			Case 1 &0.8588 dB & 1.0922 dB \\
			\hline
			Case 2 & 0.5558 dB & 0.9138 dB \\
			\hline
		\end{tabular}
		\label{tab:recepower}
		\vspace{-0.4 cm}
	\end{table}
	
\begin{figure}[t]
		\vspace{-0.5cm}
	\centering
	\subfloat[Scenario without obstructed vehicles]{\includegraphics[width=0.45\textwidth]{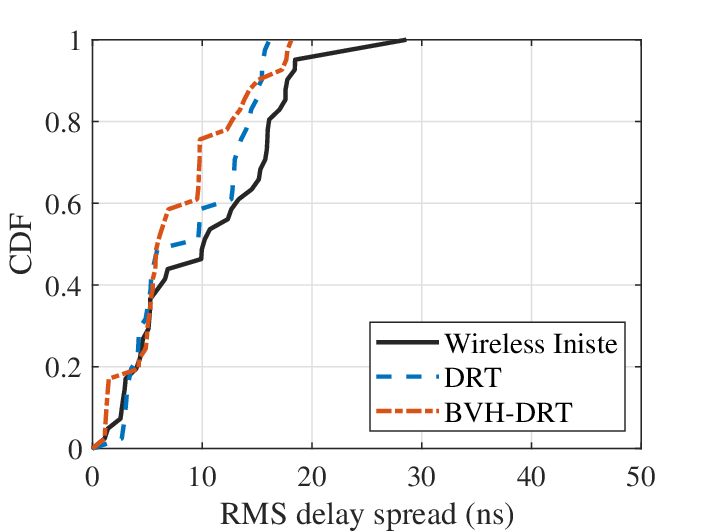}}
	\hfil
		\subfloat[Scenario with obstructed vehicles ]{\includegraphics[width=0.45\textwidth]{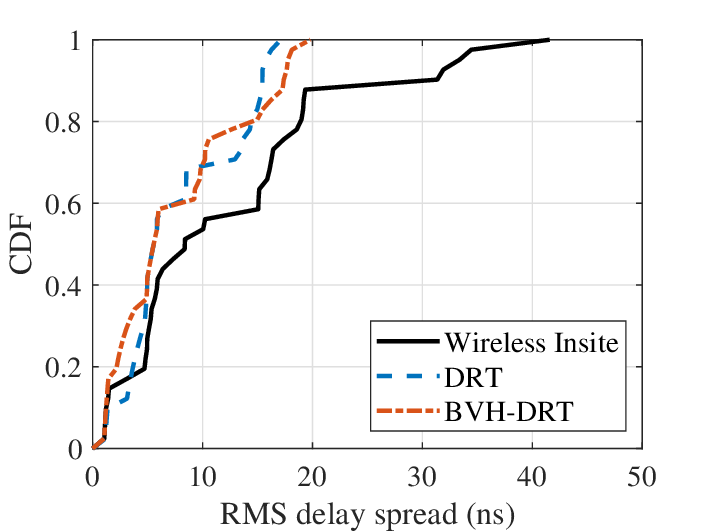}}
	\hfil
	\caption{Comparisons of RMS delay spreads among Wireless Insite, DRT, and BVH-DRT with/without obstructed vehicles: (a) scenario without obstructed vehicles and (b) scenario with obstructed vehicles.}
	\label{fig:RMSdealy}
	\vspace{-0.5cm}
	\end{figure}

	Fig. \ref{fig:RMSdealy} illustrates the RMS delay spreads over a 4 s observation window, comparing Wireless Insite, DRT, and the proposed BVH-DRT. 
	In Fig. \ref{fig:RMSdealy}(a), the RMS delay spreads computed by both DRT and BVH-DRT converge at approximately 20 ns, demonstrating high consistency. 
	The results obtained from Wireless InSite exhibit slightly larger values, as the commercial software captures a higher density of multipath components without the simplifications applied in the DRT-based algorithms. 
	Overall, the strong agreement among the three methods effectively validates the accuracy of the proposed BVH-DRT. 
	Moreover, Fig. \ref{fig:RMSdealy}(b) explores the impact of dynamic obstructed vehicles on the RMS delay spread. 
	A cross-case comparison reveals a noticeable increase in the RMS delay spread when moving vehicles exist because the dynamic blockages increase multipath richness. 
	Even within this more complex environment, the BVH-DRT remains highly consistent with both DRT and Wireless Insite.

\begin{figure}[t]
		\centering
		\subfloat[Scenario without obstructed vehicles]{\includegraphics[width=0.45\textwidth]{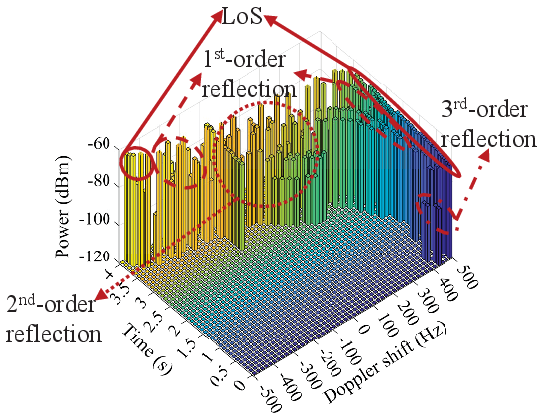}}
		\hfil
		\subfloat[Scenario with obstructed vehicles ]{\includegraphics[width=0.45\textwidth]{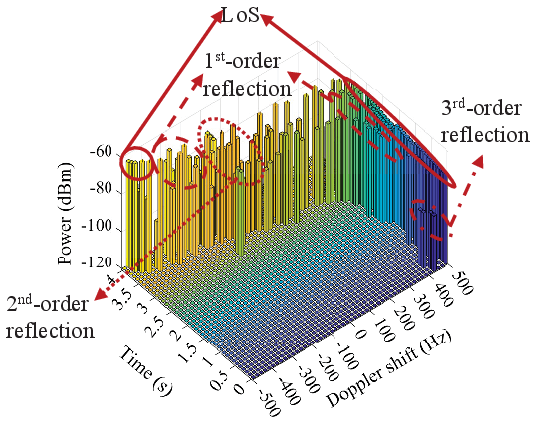}}
		\hfil
		\caption{The Doppler shifts for the BVH-DRT with and without obstructed vehicles: (a) scenario without obstructed vehicles and (b) scenario with obstructed vehicles.}
		\label{fig:Doppler}
		\vspace{-0.5cm}
	\end{figure}

	Fig. \ref{fig:Doppler}(a) shows the Doppler power spectral density (PSD) computed by the BVH-DRT without obstructed vehicles. 
	The main components of Doppler PSD are marked, including LoS, first-order reflection, second-order reflection, and third-order reflection paths. 
	These components demonstrate their stability and trends over different time periods. 
	During the simulation period, the LoS path exists from start to finish.
	At the beginning of the simulation, the LoS path exhibits a positive Doppler shift.
	However, when the Tx and Rx approach and pass each other, the frequency shift becomes negative with a significant inflection point. 
	The reflection paths from buildings and parked vehicles increase when the Tx and Rx approach together in the middle of the lane.
	These observations reveal a significant effect of the relative motion of the Tx and Rx on the Doppler effect.
	Fig. \ref{fig:Doppler}(b) shows the Doppler PSDs computed by the BVH-DRT with obstructed vehicles. 
	Compared with Case 1, the moving obstructed vehicles significantly affect the birth and death of the multipath and the dynamic change of its power distribution. 
	This further proves the adaptability and accuracy of the BVH-DRT in high-dynamic scenarios.

\begin{figure}[t]
		\centering
		\subfloat[BVH-DRT ]{\includegraphics[width=0.45\textwidth]{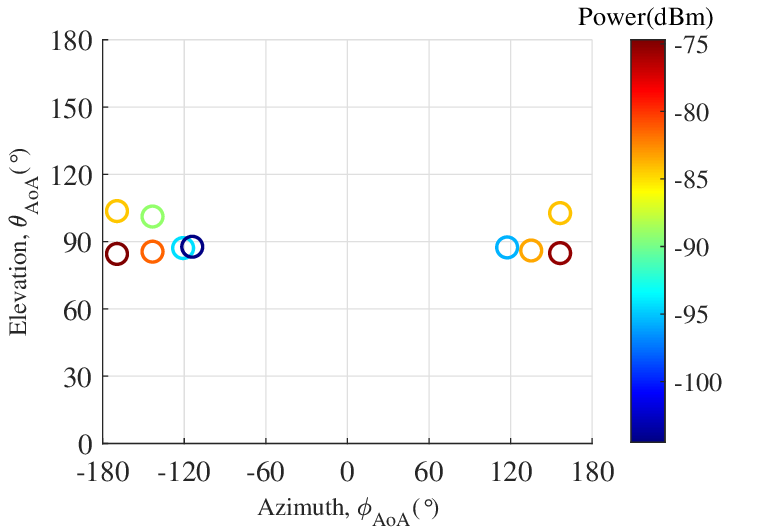}}
		\hfil
		\subfloat[Wireless Insite]{\includegraphics[width=0.45\textwidth]{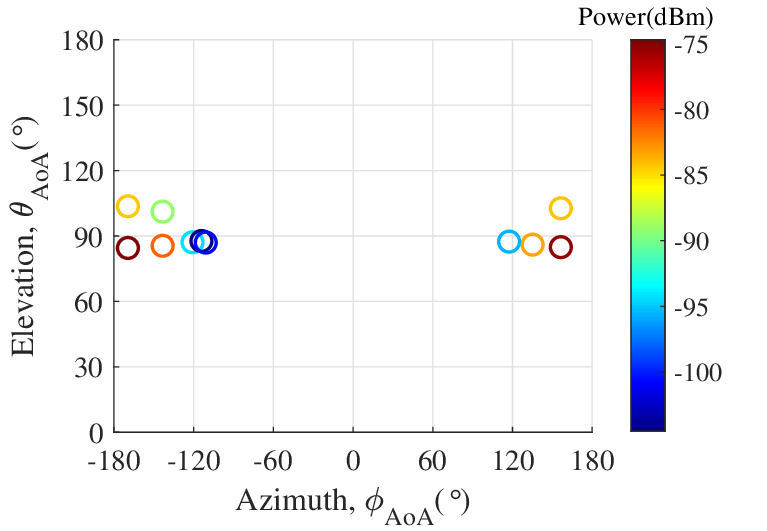}}
		\hfil
		\caption{The azimuth and elevation AoAs for the BVH-DRT and Wireless Insite without obstructed vehicles.}
		\label{fig:AoA}
		\vspace{-0.5cm}
	\end{figure}

\begin{figure}[t]
		\centering
		\subfloat[BVH-DRT ]{\includegraphics[width=0.45\textwidth]{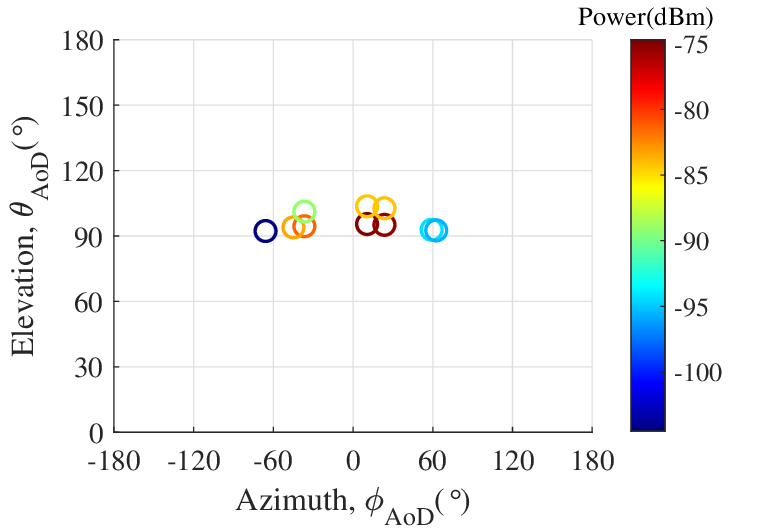}}
		\hfil
		\subfloat[Wireless Insite]{\includegraphics[width=0.45\textwidth]{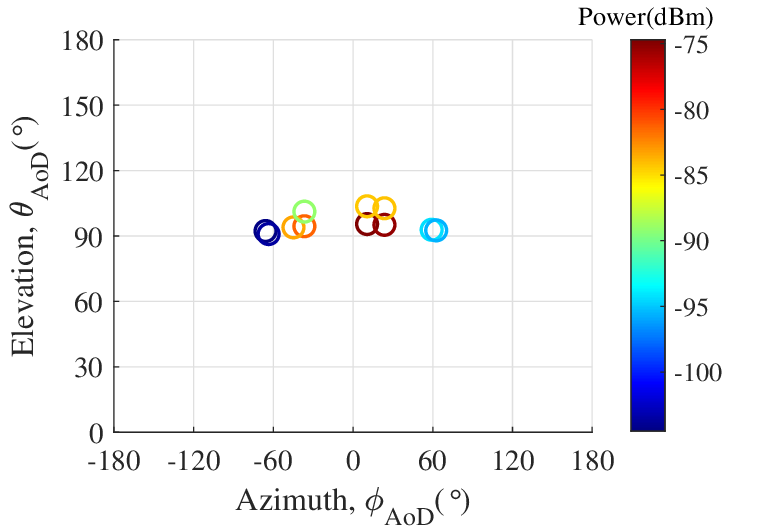}}
		\hfil
		\caption{The azimuth and elevation AoDs for the BVH-DRT and Wireless Insite without obstructed vehicles.}
		\label{fig:AoD}
		\vspace{-0.5cm}
	\end{figure}

	To further evaluate the spatial characteristics of the channel, the angular PSD of the proposed BVH-DRT and Wireless Insite at $t = 2.4$~s is analyzed in Fig.~\ref{fig:AoA} and Fig.~\ref{fig:AoD}.
	Specifically, Fig.~\ref{fig:AoA} illustrates the elevation and azimuth angle of arrival~(AoA) distribution, where the spatial profiles obtained by BVH-DRT and Wireless Insite demonstrate high consistency. 
	Additionally, the elevation AoA is highly concentrated within the $80^\circ$ to $120^\circ$ range. 
	This narrow concentration indicates a relatively simple propagation environment dominated by the LoS path. 
	In contrast, the azimuth AoA exhibits a more scattered distribution, revealing that a notable number of multipath components still exist alongside the LoS signal. 
	Similarly, Fig.~\ref{fig:AoD} presents the elevation and azimuth angle of departure (AoD) for BVH-DRT and Wireless Insite, respectively. 
	The elevation AoD aligns with the AoA trends, concentrating between $80^\circ$ and $120^\circ$, while the azimuth AoD is primarily distributed within $-50^\circ$ to $50^\circ$. 
	
\subsection{Computational Efficiency}
\begin{table}
	\centering
	\caption{Computational time for image method, DRT, and BVH-DRT with and without obstructed vehicles.}
	\renewcommand{\arraystretch}{1.2}
	\begin{tabular}{|c|c|c|c|}
		\hline
		Scenario & Image method & DRT & BVH-DRT \\
		\hline
		Case 1 & 348 s& 99 s& 85 s \\
		\hline
		Case 2 & 449 s& 114 s& 108 s \\
		\hline
	\end{tabular}
	\label{tab:Computation}
	\vspace{-0.5cm}
\end{table}

\begin{table}[t]
	\centering
	\caption{Computational time for DRT and BVH-DRT with different initial ray creation and suppression time without obstructed vehicles.}
	\renewcommand{\arraystretch}{1.2}
	\begin{tabular}{|c|c|c|c|}
		\hline
		Method &DRT &BVH-DRT & Reduction \\
		\hline
		$T_\mathrm{cs}$=0.1 s & 287s & 86 s &  \textbf{70.0\%} \\
		\hline
		$T_\mathrm{cs}$=0.2 s  & 183 s & 85 s & \textbf{53.6\%} \\
		\hline
		$T_\mathrm{cs}$=0.3 s & 99 s & 85 s& \textbf{14.1\%} \\
		\hline
		$T_\mathrm{cs}$=0.4 s & 69 s & 81 s& \textbf{-17.4\%} \\
		\hline
	\end{tabular}
	\label{tab:Timeeffi}
	\vspace{-0.5 cm}
\end{table}

\begin{table*}[t]
	\centering
	\caption{Static-BVH cluster size, complexity, and accuracy under different leaf-node thresholds.}
	\label{tab:static_bvh}
	\renewcommand{\arraystretch}{1.2}
	\begin{tabular}{|c|c|c|c|c|c|c|c|c|}
		\hline
		\multirow{2}{*}{$N_t^{\mathrm{s}}$}
		& \multicolumn{3}{c|}{Cluster size (facets per leaf)}
		& \multirow{2}{*}{Leaf count}
		& \multirow{2}{*}{Maximum depth}
		& \multirow{2}{*}{Mean build time (ms)}
		& \multirow{2}{*}{Full ray-tracing executions}
		& \multirow{2}{*}{RMSE (dB)} \\
		\cline{2-4}
		& Min. & Mean & Max. & & & & & \\
		\hline
		10
		& 2 & 5.79 & 10
		& 39 & 9 & 32.23 & 28 & 0.41 \\
		\hline
		30
		& 6 & 20.55 & 30
		& 11 & 5 & 16.40 & 21 & 0.41 \\
		\hline
		\textbf{50}
		& \textbf{19} & \textbf{32.29} & \textbf{38}
		& \textbf{7} & \textbf{4} & \textbf{11.67}
		& \textbf{13} & \textbf{0.56} \\
		\hline
		100
		& 64 & 75.33 & 87
		& 3 & 3 & 8.66 & 7 & 1.54 \\
		\hline
	\end{tabular}
\end{table*}
\begin{table*}[t]
	\centering
	\caption{Dynamic-BVH cluster size, complexity, and accuracy under different leaf-node thresholds.}
	\label{tab:dynamic_bvh}
	\renewcommand{\arraystretch}{1.2}
	\begin{tabular}{|c|c|c|c|c|c|c|c|c|}
		\hline
		\multirow{2}{*}{$N_t^{\mathrm{d}}$}
		& \multicolumn{3}{c|}{Cluster size (facets per leaf)}
		& \multirow{2}{*}{Leaf count}
		& \multirow{2}{*}{Maximum depth}
		& \multirow{2}{*}{Mean build time (ms)}
		& \multirow{2}{*}{Full ray-tracing executions}
		& \multirow{2}{*}{RMSE (dB)} \\
		\cline{2-4}
		& Min. & Mean & Max. & & & & & \\
		\hline
		1
		& 1 & 1.00 & 1
		& 12 & 7 & 0.84 & 16 & 0.54 \\
		\hline
		6
		& 1 & 3.00 & 6
		& 4 & 4 & 0.41 & 14 & 0.56 \\
		\hline
		12
		& 12 & 12.00 & 12
		& 1 & 1 & 0.02 & 13 & 0.56 \\
		\hline
		\textbf{30}
		& \textbf{12} & \textbf{12.00} & \textbf{12}
		& \textbf{1} & \textbf{1} & \textbf{0.02}
		& \textbf{13} & \textbf{0.56} \\
		\hline
	\end{tabular}
\vspace{-0.5cm}
\end{table*}

\begin{table}[t]
	\centering
	\caption{Accuracy and computational time under different reflection orders in Case 2.}
	\label{tab:reflection_order}
	\renewcommand{\arraystretch}{1.2}
	\begin{tabular}{|c|c|c|c|}
		\hline
		Reflection order
		& Method
		& RMSE (dB)
		& Time (s) \\
		\hline
		\multirow{2}{*}{1}
		& DRT & 1.07 & 23 \\
		\cline{2-4}
		& BVH-DRT & 1.01 & 19 \\
		\hline
		\multirow{2}{*}{2}
		& DRT & 0.74 & 27 \\
		\cline{2-4}
		& BVH-DRT & 0.62 & 26 \\
		\hline
		\multirow{2}{*}{3}
		& DRT & 0.73 & 114 \\
		\cline{2-4}
		& BVH-DRT & 0.56 & 108\\
		\hline
	\end{tabular}
\vspace{-0.5cm}
\end{table}
	To rigorously evaluate the computational efficiency of the proposed BVH-DRT, execution times are benchmarked against the DRT and image method. 
	All simulations are conducted on a standard desktop workstation equipped with an Intel Core i7-10700 2.9 GHz CPU and 32~GB RAM, utilizing the MATLAB platform. 
	The evaluation is divided into two main comparisons with simulation time and different initial ray creation and suppression times.
	
	Table \ref{tab:Computation} summarizes the simulation times in urban vehicular scenarios with and without moving obstructed vehicles. 
	The initial ray creation and suppression time $T_\mathrm{cs} $ is 0.3~s. 
	The results demonstrate that both DRT-based algorithms drastically reduce the computational burden compared to the image method.
	Specifically, compared to the image method, the proposed BVH-DRT achieves a remarkable computational time reduction of 75.6\% and 75.9\% in the scenarios without and with moving scatterers, respectively. 
	Furthermore, the simulation time of the BVH-DRT is slightly faster than that of the DRT. 
	This parity occurs because the adaptively adjusted ray update times in the BVH-DRT average out to a duration similar to the fixed 0.4~s setting of the DRT. 
	Consequently, as established in the accuracy evaluations, the BVH-DRT succeeds in providing superior modeling precision without incurring any additional computational overhead.

	The robust efficiency of the BVH-DRT is further highlighted when analyzing the impact of different initial ray creation and suppression times, as detailed in Table \ref{tab:Timeeffi}. 
	To capture more environmental changes and improve accuracy, the DRT requires a manually shortened fixed ray creation and suppression time. 
	However, as $T_\mathrm{cs}$ decreases, e.g., to 0.1~s or 0.2~s, the computational time of the DRT increases severely, effectively degenerating toward the complexity of the image method. 
	In contrast, the simulation time of the BVH-DRT remains remarkably stable regardless of the initial $T_\mathrm{cs}$ configuration. 
	This stability proves that the proposed adaptive adjustment mechanism intelligently regulates the ray creation and suppression cycles based on the actual environment.
	Additionally, when the $T_\mathrm{cs}$ increases, the computational time of the DRT may be less than that of the BVH-DRT because the average extrapolation interval of the DRT is larger than that of the BVH-DRT.
	However, the accuracy of the DRT decreases with the larger average extrapolation interval.
	
	The leaf-node thresholds were empirically set to $N_t^{\mathrm{s}}=50$ and $N_t^{\mathrm{d}}=30$ for the static and dynamic BVHs, respectively. These thresholds were selected to balance the spatial resolution of the leaf-node bounding boxes used for adaptive update determination against the BVH construction complexity. To evaluate the impact of leaf-node thresholds, the static and dynamic thresholds were varied separately, and the resulting cluster-size statistics, leaf counts, maximum depths, and mean construction times are summarized in Table~\ref{tab:static_bvh} and Table~\ref{tab:dynamic_bvh}. For the static BVH, reducing $N_t^{\mathrm{s}}$ from 100 to 10 decreases the mean cluster size, while increasing the leaf count, the maximum depth, and the mean construction time. Similarly, reducing $N_t^{\mathrm{d}}$ from 30 to 1 for the dynamic BVH decreases the mean cluster size and increases the leaf count, the maximum depth, as well as the mean construction time. These results confirm that a smaller $N_t$ produces finer leaf clusters and a larger BVH, thereby increasing the construction time. In Table~\ref{tab:dynamic_bvh}, $N_t^{\mathrm{d}}=12$ and $N_t^{\mathrm{d}}=30$ give the same results because the maximum number of vehicular facets is set as 12 in Case 2.

	To evaluate the accuracy--complexity tradeoff associated with the cluster size, the maximum reflection order was fixed at 3, while \(N_t^{\mathrm{s}}\) and \(N_t^{\mathrm{d}}\) were varied separately. As shown in Table~\ref{tab:static_bvh} and Table~\ref{tab:dynamic_bvh}, smaller thresholds produce finer clusters, increase the BVH construction time, and require more full ray-tracing executions. The RMSE reduction is more evident when varying the static threshold, whereas the dynamic threshold has a limited effect on accuracy in Case~2. This is because finer clustering improves accuracy only when it reduces the effective coherence distance $D_{\mathrm{c}}$. Therefore, the selected thresholds provide an empirical trade-off between accuracy and computational complexity.
	
	As shown in Table~\ref{tab:reflection_order}, the two methods exhibit different accuracy--complexity trends as the maximum reflection order increases. The RMSE of received power for the two methods is calculated against the Wireless Insite results. For DRT with a fixed $T_{\mathrm{cs}}=0.3$ s, increasing the reflection order from 1 to 2 reduces the RMSE from 1.07~dB to 0.74~dB, while increasing the simulation time from 23~s to 27~s. Increasing the order from 2 to 3 further reduces the RMSE to only 0.73~dB, but increases the simulation time to 114~s. For BVH-DRT, the RMSE decreases from 1.01~dB to 0.62~dB and 0.56~dB as the reflection order increases, while the corresponding simulation time increases from 19~s to 26~s and 108~s. At each tested reflection order, BVH-DRT achieves both a lower RMSE and a shorter simulation time than DRT. Overall, the second-order reflection provides an accuracy--complexity balance in Case~2.

\subsection{BER Performance}
	\begin{figure}[t]
	\centering
	\includegraphics[width=0.85\linewidth]{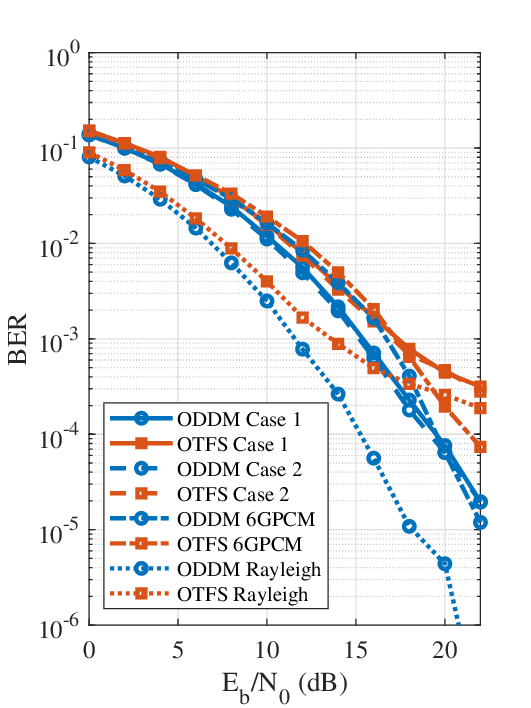}
	\caption{BER performances of ODDM and OTFS under the BVH-DRT channels for Cases 1 and 2, the Rayleigh fading channel, and the 6GPCM~\cite{CXWang2022TVT}.}
	\label{fig:BER}
	\vspace{-0.5cm}
\end{figure}
	Fig.~\ref{fig:BER} illustrates the bit error rate (BER) performance of ODDM and OTFS under the BVH-DRT channels for Cases 1 and 2, the Rayleigh fading baseline, and the 6GPCM benchmark~\cite{CXWang2022TVT}. To generate the GBSM benchmark, 6GPCM is configured using the 3GPP 37.885 Urban non-line-of-sight vehicle‌ (NLOSv) scenario~\cite{3GPP2019Study}. The carrier frequency, Tx and Rx trajectories, and simulation duration are consistent with the BVH-DRT evaluation. The same waveform parameters, per-snapshot channel-power normalization, channel state information, and LMMSE equalizer are applied to all channel models. For the two BVH-DRT cases, the Rayleigh baseline and the 6GPCM, ODDM consistently achieves a lower BER than OTFS. In Case 2, the performance gap between the two waveforms becomes more pronounced because the bi-orthogonal pulse-shaping mismatch of OTFS increases the inter-symbol interference and inter-carrier interference under severe time-frequency dispersion, whereas ODDM retains strict orthogonality in the DD domain. The BER values vary among the evaluated channel models because 6GPCM provides a statistically generated time-evolving channel, whereas BVH-DRT resolves the site-specific propagation paths determined by the considered street geometry and moving obstructed vehicle. Thus, the 6GPCM comparison confirms the consistent relative waveform behavior, while BVH-DRT additionally enables the evaluation of geometry-dependent and obstruction-dependent channel variation.

\section{Conclusions}
	In this paper, we have proposed the BVH-DRT framework to overcome the computational and accuracy bottlenecks in DRT	by redefining the update mechanism from an unpredictable coherence time to a spatial-driven coherence distance.
	The integration of BVH as a pre-processing step has clustered complex static and dynamic scatterers, enabling the adaptive adjustment of ray creation and suppression time based on transceiver kinematics. 
	Validation against channel measurements and commercial software in a complex urban V2V scenario has confirmed that BVH-DRT has accurately captured critical fading characteristics. 
	Furthermore, BVH-DRT has significantly reduced the overall computational burden. 
	BER evaluations for OTFS and ODDM systems have confirmed that the BVH-DRT framework has precisely reflected the severe performance degradation induced by complex dynamic blockages. 
	Ultimately, the BVH-DRT framework has provided a highly efficient and accurate channel modeling tool, satisfying the strict demands of 6G high-dynamic wireless networks.

\end{document}